\documentclass[aip,jcp,reprint]{revtex4-1}

\pdfoutput=1
\usepackage[utf8]{inputenc}
\usepackage[version=3]{mhchem}
\usepackage{graphicx}
\usepackage{hyperref}

\def\AR{\texttt{AR} }
\def\ttA{\textrm{A}}
\def\ttT{\textrm{T}}

\def\fbc{\texttt{fix bond/create}}

\begin{document}

\title{A parallel algorithm for step- and chain-growth polymerization in Molecular Dynamics}
\author{Pierre de Buyl}
\affiliation{Division of Polymer Chemistry, Department of Chemistry, Katholieke Universiteit Leuven, Celestijnenlaan 200F, B-3001 Heverlee, Belgium}
\affiliation{SIM vzw, Technologiepark 935,BE-9052 Zwijnaarde, Belgium}
\author{Erik Nies}
\affiliation{Division of Polymer Chemistry, Department of Chemistry, Katholieke Universiteit Leuven, Celestijnenlaan 200F, B-3001 Heverlee, Belgium}
\begin{abstract}
  Classical Molecular Dynamics (MD) simulations provide insight on the
  properties of many soft-matter systems.
  In some situations it is interesting to model the creation of chemical bonds,
  a process that is not part of the MD framework.
  In this context, we propose a parallel algorithm for step- and chain-growth
  polymerization that is based on a generic reaction scheme, works at a given
  intrinsic rate and produces continuous trajectories.
  We present an implementation in the ESPResSo++ simulation software and compare
  it with the corresponding feature in LAMMPS.
  For chain growth, our results are compared to the existing simulation literature.
  For step growth, a rate equation is proposed for the evolution of the
  crosslinker population that compares well to the simulations for low
  crosslinker functionality or for short times.
\end{abstract}
\maketitle

\section{Introduction}

For many applications in soft-matter research, chemical bonds can be considered
a given data that does not change in the course of time.
For instance, the chemical structure of water is typically not modified in a
molecular simulation when it is used as a solvent~\cite{frenkel_smit_2001}. Likewise, the
structure and chemical bonds of complex molecules are typically fixed in the
course of a simulation.

Creating new chemical bonds in a molecular simulation is a problem for which no
general solution exists. This is due to the inherent complexity of the
problem at hand, as chemical reactions are not part of the Hamiltonian mechanics
paradigm that serves as the basis for classical Molecular Dynamics (MD) simulations.
Ab initio simulations could, in principle, be used for this purpose but the
computational cost remains prohibitive.
As a result, several approaches have been followed in the literature to model
chemical bonding in MD. Farah {\em et al}~\cite{farah_et_al_cpc_2012} classify
the reaction methods between empirical force-fields and methods based on a
reaction cutoff distance.
In the cutoff method, one adds a bonded interaction between particles based on
their type and their distance. This approach has been applied to coarse-grained
models~\cite{akkermans_et_al_jcp_1998,stevens_connectivity_2001,hoy_fredrickson_jcp_2009}
and to atomistic
models~\cite{heine_et_al_macromolecules_2004,wu_xu_polymer_2006,varshney_et_al_macromolecules_2008},
using different simulation protocols. In general, this approach relies on a
cutoff distance that is larger than the typical interaction between particles,
leading to artificially large bond distances upon binding, %
energy jumps and discontinuous trajectories. %
These issues are typically silenced by the use of thermostatting. %
The polymerization protocol proposed by Akkermans {\em et
  al}~\cite{akkermans_et_al_jcp_1998} prevents the discontinuities and allows to
simulate properly thermoneutral reactions. Endo- and exo-thermic reactions are
not considered here.
Another typical feature of the cutoff protocols is that the binding is applied
between MD runs via external single-CPU programs, which imposes to keep the
number of bindings steps reasonable.
The MD code LAMMPS~\cite{plimpton_lammps_1995} offers a bond creation feature,
\texttt{fix bond/create}, that works in parallel and during the simulation,
allowing a continuous application of the reaction step. %
This feature, although its use appears in the literature (see for instance
Ref.~\onlinecite{odegard_2014} where it is used to prepare a system for ReaxFF),
has not been the topic of a dedicated publication and, more specifically, its
kinetic properties have not been studied. %
We review its implementation in Sec.~\ref{sec:existing} and compare its
polymerization kinetics with our algorithm.

Empirical force-fields (see Ref.~\onlinecite{farah_et_al_cpc_2012} and
references therein) have been designed to model bond formation and breaking
in MD and aim at reproducing a continuous transition of the chemical bonds from
unbonded to bonded particles. %
ReaxFF~\cite{van_duin_et_al_reaxff_jpca_2001} is such an empirical force-field,
it builds on ab initio data to reproduce the interactions in a dynamical
approach: the parameters for the interatomic force fields are updated at each
step to resemble those of a full quantum simulation.
ReaxFF brings a great level of detail at a lesser cost than a full quantum
simulation but does not allow yet to simulate systems as large a classical MD
allows.
The use of cutoff methods, such as the one presented here, remains of great
importance either to study generic aspects of polymerization or as a way to
prepare configurations for further atomic simulations, as is done in
Ref.~\onlinecite{odegard_2014}.

In the present article, we focus on coarse grained models for the simulation of
polymer systems. Their simplicity, in comparison to atomistic models, allows us
to devise a consistent polymerization procedure. We consider only distance-dependent
pairwise interactions between the particles that participate in the chemical
reaction.
The algorithm is exposed in full details and is implemented in the ESPResSo++
soft-matter simulation software\cite{halverson_et_al_espressopp_cpc_2013} as an
extension that is distributed as part of the version 1.9.
The execution of the algorithm makes use of the Message Passing Interface for
distributed memory parallel computing, which ESPResSo++ already uses.
The communication pattern that is needed to perform a random partner selection
is a constitutive part of our algorithm.
The corresponding feature of LAMMPS, \fbc{}, is described on the basis of its
source code.
The algorithm developed for ESPResSo++ is then implemented in LAMMPS to address
the difference that is observed between the algorithms.
The complete simulation input (datafiles, scripts and programs) to reproduce the
results presented here is made available under the BSD license~\cite{cg_polym} for both the
LAMMPS and the ESPResSo++ implementations, insuring that no details of the
simulation protocol remains undefined.

The algorithm manages chain- and step-growth mechanisms. Rate equations and
simulations for both situations are presented and used to assess the
time-evolution of the crosslinking process.
This comparison lays a formal basis for future simulations of bond-forming
systems in a way that embeds the computational and theoretical approaches.

\section{Bond formation model and algorithm}
\label{sec:algo}

A single bonding algorithm, set with appropriate parameters, can be used to
generate different growth mechanisms.
We describe this general algorithm and its
application to the step- and chain-growth mechanisms for polymerization.
We introduce a state variable for all particles that will define
their active and/or available status. It is denoted by
\begin{equation}
  \cee{A^{s\ast}} ~,
\end{equation}
where $s$ is the {\em state} of the particle of {\em type} A.

The creation of bonds is ruled by the chemical equation
\begin{equation}
  \label{eq:general}
  \cee{T^{t\ast} + A^{a\ast} ->[k] T^{(t+\delta t)\ast}-A^{(a+\delta a)\ast}}
\end{equation}
that represents the behaviour of a {\em target} T and an {\em active} particle A.
Only irreversible reactions are considered here.
Before the reaction, T and A have no specific connection besides the nonbonded excluded volume interaction. T and A particles
in a given interaction range will react at a rate $k$, if they meet criteria
detailed in the following of the text, to form a connected compound T-A. The
state of T and A is then updated according to the reaction parameters $\delta$t
and $\delta$a.

In order to avoid discontinuities in the trajectories or in the energy of the
simulated system, the bonded interaction must not impact the particle at the time of
the reaction.
This is achieved by using a bonded potential that is equal to zero below a given
cut-off following the proposition by Akkermans {\em et al}~\cite{akkermans_et_al_jcp_1998}.
A consequence is that the reaction is thermoneutral and that thermostatting is
not required to absorb energy jumps as in previous studies~\cite{hoy_fredrickson_jcp_2009,mukherji_abrams_2009}.
To the authors' knowledge, controlled exo- or endothermic reaction schemes for
MD do not exist.
The interactions for other particles that do not undergo polymerization,
i.e. any particle except A and T, may be more complex however, as they will not
be impacted by the addition of the A-T bond.
This way, molecules with angular, dihedral and/or improper interactions may take
part in the polymerization.

At variance with Akkermans {\em et al}~\cite{akkermans_et_al_jcp_1998}, however, the rate of the reaction
is not controlled by the interval at which the reaction is performed ($\tau_r$
in Ref.~\onlinecite{akkermans_et_al_jcp_1998}). Instead, it is the value of $k$ that dictates the dynamics.
Reactions are performed every $\Theta$ MD steps of timestep $\Delta t$
(see Sec.~\ref{sec:implementation}) and the parameters must obey
\begin{equation}
  \label{eq:parameters}
  k\, \Delta t\, \Theta \ll 1 ~.
\end{equation}
A pair is considered for reaction if
\begin{equation}
  \label{eq:acceptance}
  u<k\, \Delta t\, \Theta~,
\end{equation}
where $u$ is a random number distributed uniformly in $[0,1)$.

\subsection{Chain growth}

Chain growth is considered to occur at the single active end of a polymer chain
of $n$ units
\begin{equation}
  \cee{P_{n-1}-P^{\ast} + M ->[k] P_n-P^{\ast}}
\end{equation}
so that the monomer M becomes the last, and active, unit of the polymer chain.
Here, P$^\ast$ is the {\em active} particle and M the {\em target} particle.
While there is a single active unit in a polymer chain, there may be many
polymer chains in a single simulation.
Every unit, except the first and last in a chain, may form two bonds: one as the
target and one as the active particle. The monomeric units are thus of functionality two.

The kinetic evolution of the population of chains is given, following Akkermans
{\it et al}~\cite{akkermans_et_al_jcp_1998} by
\begin{subequations}
  \label{eq:chain-kin}
  \begin{align}
    \dot{[P_1]} &= -k_{c} [M][P_1],\\
    \dot{[P_{n+1}]} &= -k_{c} [M] \left( [P_{n+1}] - [P_{n}] \right),
  \end{align}
\end{subequations}
where the dot denotes the time derivative.
The effective rate constant for the chains $k_{c}=k \rho^{-1} \langle N_{P^\ast M}
\rangle$ takes into account the intrinsic rate $k$ and the average number
$\langle N_{P^\ast M} \rangle$ of available monomers M around a polymer
end-unit P$^\ast$.
Following Ref.~\onlinecite{akkermans_et_al_jcp_1998}, $\langle N_{P^\ast M} \rangle$ is considered independent of
the chain length and is obtained from simulations in which the polymerization is
stopped at different lengths.
$[\cdot]$ stands for the number density (or concentration) of a given
particle type and has the units of an inverse volume.

As a consequence of Eqs.~(\ref{eq:chain-kin}), the average concentration of
monomers $[M]$ follows a simple evolution:
\begin{align}
  \dot{[M]} &= -k_{c} [M] \sum_{n=1}^{\infty} [P_{n}]\\
  &= -k_{c} [M] [P^\ast]
\end{align}
where $[P^\ast]$ is the concentration of active end-units, which is a constant
here.
The resulting concentration of monomers thus follows an exponential decay
\begin{equation}
  [M](t) = [M_0] e^{-k_{c} [P^\ast] t} ~,
\end{equation}
where $[M_0]$ is the initial value of the concentration $[M]$.
Alternatively, we may consider the polymer fraction
\begin{align}
  \label{eq:chain-fraction}
  \phi(t)&=1-\frac{[M](t)}{[M]+[P]+[P^\ast]}\cr
  &= 1 - \frac{[M_0]}{[M]+[P]+[P]^\ast} e^{-k_{c} [P^\ast] t}\cr
  &\approx 1 - e^{-k_{c} [P^\ast] t} ~,
\end{align}
where the approximation accounts for the fact that almost all particles in the
system are available monomers at the beginning of the simulation.
There is no termination in the algorithm: polymerization stops only when the
program finds no further candidate pairs, either because the system is depleted
in available monomers M or because the available monomers M are not in the
vicinity of active units P$^\ast$.

\subsection{Step growth}

Step growth is considered here in the case of a crosslinker X joining
E-P$_{n}$-E chains, where E stands for ``end unit'' and there are $n$ repeat
units within the chain. We consider only the reaction of the crosslinker at the
end units.
This is a representative situation for epoxy materials for instance, in which X
is also called the {\em curing agent}, and is typical of step
growth~\cite{stevens_polychem_1999}.

The reaction mechanism is
\begin{equation}
  \cee{E-P_n-E^{0\ast} + X^{s\ast} ->[k] E-P_n-E^{1\ast}-X^{(s-1)\ast}}
\end{equation}
in which the state of the left-end unit E is not relevant, it may be either free
or already linked to a crosslinker.
Here, X$^{s\ast}$ is the {\em active} particle and $E^{0\ast}$ the {\em target} particle.
The crosslinker may have other bonds already, as long as $s>0$.
The crosslinkers are given an initial state $s_0=f$ that corresponds to their
chemical functionality $f$: the algorithm lets them form bonds up to $f$ times.
When $s$ reaches $0$, the algorithm stops the formation of further bonds.

An approximation for the kinetic evolution of the concentration of state $s$
crosslinkers is
\begin{subequations}
  \label{eq:step-kin}
  \begin{align}
    \dot{[X_0]} &= -k_0 [X_0],\\
    \dot{[X_s]} &= -k_s [X_s] + k_{s-1} [X_{s-1}],\label{eq:step-kin-s}\\
    \dot{[X_f]} &= k_{f-1} [X_{f-1}],
  \end{align}
\end{subequations}
where Eq.~(\ref{eq:step-kin-s}) is
valid for $0<s<f$.
$k_s$ is the effective reaction rate that depends on $k$ and on $\langle
N_{X^sE^0} \rangle$
\begin{equation}
  k_s = \langle N_{X^sE^0} \rangle k ~,
\end{equation}
where $\langle N_{X^sE^0} \rangle$ is the number of potential partners that may
enter reaction \eqref{eq:step-kin}; it will be determined by the radial
distribution function later on.
Results will be displayed with the number fractions of crosslinkers
\begin{equation}
  \label{eq:step-fraction}
  x_s = \frac{[X_{s}]}{\sum_{s'=0}^f [X_{s'}]} ~.
\end{equation}
Equation~\eqref{eq:step-kin} is solved numerically with the routine {\tt odeint}
from SciPy~\cite{scipy-web} {\tt integrate} module, using $x_0=1$ and $x_s=0$ for the initial
value.

The rate equation \eqref{eq:step-kin} provides a comparison point for
the simulations, with the following limitations:
(i) the equation neglects correlations in the system and
(ii) it accounts for the structure only via the average values for $\langle
N_{X^sE^0} \rangle$.
The role of $\langle N_{X^sE^0} \rangle$ in the rate equation is to reproduce
the steric hindrance around a crosslinker X: if X is already connected to $f-s$
E particles, there is a corresponding lack of space for further E particles to
connect to X.

\subsection{Simulation details}

To complete the bond formation model, we present here the Molecular Dynamics
(MD) configuration with which the simulations of sections \ref{sec:sim-chain}
and \ref{sec:sim-step} have been performed.
All simulations are run using either
ESPResSo++\cite{halverson_et_al_espressopp_cpc_2013} version 1.9 or
LAMMPS~\cite{plimpton_lammps_1995} (with the source code for the new
algorithm~\cite{fbc_random}).

All the particles in the system have identical masses $m$ and interact via a
truncated Lennard-Jones 6-12 potential
\begin{align}
  \label{eq:LJ}
  V_{LJ}(r) &= 4 \epsilon \left( \left(\frac{\sigma}{r}\right)^{12} - \left(\frac{\sigma}{r}\right)^{6} + \frac{1}{4} \right) \cr
  &\phantom{= 0 \quad} \textrm{for } r<\sigma_c ~,\cr
  &= 0 \quad \textrm{else.}
\end{align}
The $\epsilon$ and $\sigma$ parameters are the same for all monomer and
crosslinker particles. All quantities are reported in reduced Lennard-Jones
units of mass $m$, length $\sigma$, energy $\epsilon$ and time $\sigma\sqrt{{m}{\epsilon^{-1}}}$~.

Polymer chains in the step-growth simulations are held together by a FENE
potential
\begin{equation}
  V_{F}(r) = -\frac{1}{2} k_{F} R_0^2 \ln\left(1 - \left(\frac{r}{R_0}\right)^2\right)
\end{equation}
using the Kremer-Grest~\cite{kremer_grest_jcp_1990} parameters $k_{F}=30$ and $R_0=1.5$~.

The bonds that are created during the simulations are modeled with a mirror
Lennard-Jones potential~\cite{akkermans_et_al_jcp_1998} with the same parameters as in Eq.~\eqref{eq:LJ}:
\begin{align}
  \label{eq:mirrorLJ}
  V_{b}(r) &= 4 \epsilon \left( \left(\frac{\sigma}{2\sigma_c - r}\right)^{12} - \left(\frac{\sigma}{2\sigma_c - r}\right)^{6} + \frac{1}{4} \right)\cr
  &\phantom{= 0 \quad} \textrm{for } \sigma_c<r<2\sigma_c ~,\cr
  &= 0 \quad \textrm{else.}
\end{align}

A velocity-Verlet integration with timestep $\Delta t=0.0025$ is used for all
simulations. A thermostat is used to prepare the systems at
temperature $T=1$. The number density is $\rho=0.8$.
The thermostat is only used for the thermalisation of the system and is not
necessary during the polymerisation part, due the the energy conservation
property of the curing algorithm.
The explicit protocols are given in Appendix~\ref{sec:protocols} and are
available online~\cite{cg_polym}.

\section{Implementation in ESPResSo++}
\label{sec:implementation}

\begin{table*}[t!]
  \caption{\label{tab:steps}Steps performed by the \texttt{AssociationReaction}
    extension. The lists used for storing the pairs and their contents at each
    steps are given. The presence of a $\Pi$ indicates a parallel communication step.
  }
  \begin{ruledtabular}
    \begin{tabular}{lp{.4\linewidth}lp{.4\linewidth}}
      Step & Action & List & Content\\
      1. & Find all suitable candidate for a bond formation in the neighbor list of each
      particle.\\
      2. & Retain the candidates on the basis of a given rate, by comparing to a
      random number (see Eq.~\eqref{eq:acceptance}).\\
      3. & Collect, for each A particle, the list of candidate targets T. & $L_\ttA$ & Local (id$_\ttA$,id$_\ttT$) pairs, ordered by id$_\ttA$\\
      4. $\Pi$ & Consolidate the list among neighboring CPUs.&$L_\ttA$ & Local and neighboring (id$_\ttA$,id$_\ttT$) pairs, ordered by id$_\ttA$\\
      5. & Keep only one candidate pair per A particle & $L_\ttA$ & Unique (id$_\ttA$,id$_\ttT$) pairs, with respect to id$_\ttA$\\
      6. $\Pi$ & Assemble the candidate list for the target particles only. & $L_\ttT$ & Local and neighboring (id$_\ttT$,id$_\ttA$) pairs, ordered by id$_\ttT$\\
      7. & Select randomly, for each target particle, one activated particle.&$L_\ttT$ & Unique (id$_\ttT$,id$_\ttA$) pairs, with respect to id$_\ttT$\\
      8. $\Pi$ & Communicate the selected pairs among neighboring CPUs.\\
      9. & For all the selected pairs: add the bond, modify the states for the active and the target particle.\\
    \end{tabular}
  \end{ruledtabular}
\end{table*}

\begin{figure}[h]
  \centering
  \includegraphics[width=.75\linewidth]{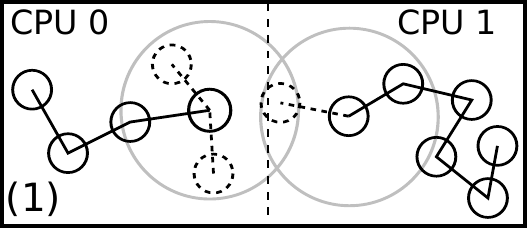}
  \includegraphics[width=.75\linewidth]{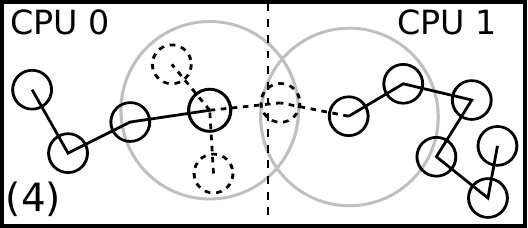}
  \includegraphics[width=.75\linewidth]{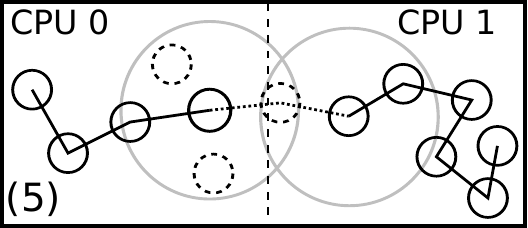}
  \includegraphics[width=.75\linewidth]{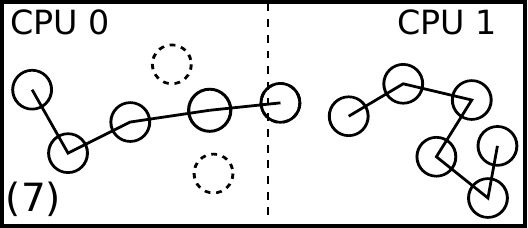}
  \caption{Illustration of the picking algorithm for a situation where a target
    particle is selected by two active particles. The step numbers are from
    table~\ref{tab:steps}.
    The gray circles indicate the reaction cutoff
    for the two active particles. From top to bottom:
    (1) The active particle (one on CPU 0 and one on CPU 1) finds all of its
    partners (potential partners are dashed and connected with dashed lines).
    (4) The partners from neighboring CPUs are collected.
    (5) Each active particle selects one partner (here, the same target particle
    is selected on CPU 0 and CPU 1). This selection is shown with a dotted line.
    (7) The target particles only allows a single particle to make the reaction.
  }
  \label{fig:steps-cpus}
\end{figure}

The algorithm presented in Sec.~\ref{sec:algo} has been implemented in
ESPResSo++~\cite{halverson_et_al_espressopp_cpc_2013}. ESPResSo++ is designed for extensibility at two levels: (i)
the software is used by writing Python programs, in which it is possible to
interact in a powerful manner with the system that is simulated and (ii) C++
extensions can be ``plugged in'' to modify the execution of the main simulation
algorithms at many places.
It is possible to add arbitrary operations at specific positions within the main MD looop.

We have written an extension, \texttt{AssociationReaction} (or \texttt{AR} for
short) that is executed within the Molecular Dynamics integrator, after the
Velocity Verlet and thermostat have been run, i.e. it is connected to the
\texttt{aftIntV} signal.
The algorithm is run every $\Theta$ time steps. The specific value of $\Theta$
does not affect the result as it is taken into account in the acceptance
criterion~\eqref{eq:acceptance}.

The behavior of the \AR extension is influenced by the following parameters:
the types of T and A, the
minimum state for A $s_{min}^A$, $\delta$T, $\delta$A, the rate $k$, the interval
$\Theta$ at which \AR is run and the cutoff for the reaction.

\subsection{Parallel communication}

In order to work in parallel, information on the bond choices must be
communicated among neighboring processors.
An important component of the implementation is the routine
\texttt{sendMultiMap} that consolidates the candidate lists among neighboring
CPUs and that is used three times for a reaction step (at each symbol $\Pi$ in
table~\ref{tab:steps}).

For the sake of clarity, the use of several terms is given in the context of
parallel programming:
\begin{description}
\item[CPU] A processing unit that acts as a MPI worker.
\item[neighboring CPU] A CPU that is in direct contact with a given one. Each
  CPU has 8 neighboring CPUs.
\item[ghost] A ghost is a particle whose data is present on a CPU but for which
  the equations of motion are not solved. The presence of ghosts is necessary to
  compute force or reaction decisions.
\end{description}

We give here the explicit sequence of steps that are run by \AR. The neighbors
pairs are taken from the existing Verlet list that is used for the Lennard-Jones
interaction.
This convenience is possible because we select a cutoff for the reaction scheme
that is the same as for the Lennard-Jones interaction.
The communication pattern follows the implementation in
\texttt{storage::DomainDecomposition}.
\begin{enumerate}
\item For all neighbor pairs, collect the ones matching the type and state given
  as parameters as pairs (id$_\textrm{A}$, id$_\textrm{T}$).
\item Retain the matching pairs with rate $k$.
\item On each CPU, collect the pairs sorted by their $id_\textrm{A}$ value in
  the list $L_\textrm{A}$.
\item Communicate $L_\textrm{A}$ to all neighboring CPUs and merge the local and
  adjacent $L_\textrm{A}$.
\item On the basis of $L_\textrm{A}$, select only one T particle, at random, to
  react with each A. This choice is made on the CPU for which A is not a ghost.
\item Communicate $L_\textrm{A}$ to all neighboring CPUs and merge the local and
  adjacent $L_\textrm{A}$.
\item On the basis of $L_\textrm{A}$, select only one A particle, at random, to
  react with each T, and collect them in the list $L_\textrm{T}$.
\item Communicate $L_\textrm{T}$ to all neighboring CPUs and merge the local and
  adjacent $L_\textrm{T}$.
\item Apply the change to the states of the A and T particles from $L_\textrm{T}$.
\end{enumerate}

The algorithm is detailed in table~\ref{tab:steps} with the explicit mention of
the communication steps and the content of the pair lists.

This algorithm ensures that each A or T particle can only participate in one new
bond at each time step, even if several candidate bonds exist.
This is achieved by selecting successively the pairs for a unique A and {\em
  also} for a unique T from the local {\em and} neighboring CPUs.
This problem is illustrated in Fig.~\ref{fig:steps-cpus}.
The overall reaction rate depends naturally on the number of neighbor A T pairs
in the system.

The effective bond formation is implemented by adding a bonded interaction term
in the MD simulation.
Explicitly, this amounts to call the {\tt add} methods on the {\tt
  FixedPairList} that contains the bonded Mirror Lennard-Jones interaction
potential.

\subsection{Current limitations}

There are several possible extensions to the algorithm that would bring more
generality.
Taking into account several concurrent reactions is possible, following the
Reactive multiparticle collision dynamics algorithm presented in Rohlf {\it et
  al}~\cite{rohlf_et_al_rmpcd_2008} for collision-based hydrodynamical
simulations.
Further, the algorithm only considers irreversible reactions. Adding
dissociation reactions would require an interaction potential that can be cut
off without discontinuity. Quartic bonds have already been used for this purpose
by Panico {\em et al}~\cite{tsige_stevens_macromolecules_2004,panico_et_al_failure_2010} in the
LAMMPS~\cite{plimpton_lammps_1995} Molecular Dynamics simulation code.

\section{Implementation in LAMMPS}

\subsection{Existing implementation}
\label{sec:existing}

LAMMPS provides officially the feature \texttt{fix bond/create} since january
2009~\footnote{\url{http://lammps.sandia.gov/history.html}}, although it may
have been developed earlier as the use of the nearest partners is
mentioned in Ref.~\onlinecite{heine_et_al_macromolecules_2004}.
As the details of the implementation of \texttt{fix bond/create} in LAMMPS have
not been described in the literature, we review them here from the analysis of
the file \texttt{fix\_bond\_create.cpp}.
This \texttt{fix} operates at the \texttt{post\_integrate} step in the MD
integrator.
LAMMPS does not possess a variable {\em state} for the particles. When a change
is needed, it is done by modifying a particle's {\em type} instead.

The parameters given to the \fbc{} command are: the types of the particles A
and B, the cutoff distance for the bond creation, the bond type to create and
optionally the maximum number of bonds to create for A and B, the type in
which to transform A and B when reaching this maximum, a probability for the
bond creation and the types of the angular and dihedral interactions to create.

As in the ESPResSo++ implementation, the algorithm relies on the existing
neighbor list that is used for the non-bonded interactions.
\begin{enumerate}
\item For each neighbor pair that matches the types:
  \begin{enumerate}
  \item Test for the correspondance of the types and the cutoff criterion.
  \item If the distance of the pair is lower than the minimum that was found
    previously, record the particles' indices and distance.
  \end{enumerate}
\item The candidate pairs are consolidated among the processors using again the
closest match in distance.
\item In each of the selected pairs, the evaluation of the reaction probability
  is done on the particle with the lowest identifier (the \texttt{tag} in LAMMPS).
  A random number in $[0,1)$ is compared to the user-defined probability.
\end{enumerate}

This last criterion allows the choice of partners to be made uniquely in a
simpler process than the one presented for ESPResSo++.
The implication is that the choice of partners is not done at random among all
possible partners.
Parallel communications occur for the collection of partners and the
synchronization of the random number assigned to each partner pair. A final
communication ensures that the bond creation and type update is performed on
each CPU.

After the bonds have been created, LAMMPS updates the connectivity of the system
and checks for the generation of the angular and dihedral interactions that
could result from the new molecular bonds, if the user has requested these in
the \fbc{} instruction.
Discontinuities in these interactions will perturb the trajectory and the energy
of the system if enabled but remain a powerful feature to build atomistic
networks.

The following considerations have to be considered when using \fbc{}.
The user has to request the provision for extra connectivity information
(i.e. allocation of appropriate storage for bonds, angles and dihedras, via the
\texttt{extra bond per atom} and \texttt{extra special per atom} settings).
We have included the repulsive Lennard-Jones potential, normally part of the
nonbonded interactions, in the mirror Lennard-Jones bonded potential to follow
the behaviour of the FENE bonds in LAMMPS.
This is needed as bonded particles are excluded from the force evaluation, and
this cannot be changed when the FENE potential is in use, which is the case
here.
The \fbc{} command keeps in memory the total number of bonds created during the
simulation. If the user wishes to obtain further information on the bonds,
e.g. on their distribution, it must be obtained via a dump to disk of the
property \texttt{nbond}.

\subsection{New implementation}

As will be seen in Sec.~\ref{sec:sim-chain}, the existing algorithm in LAMMPS
produces a different polymerization kinetics than the one we designed. %
To confirm that the difference originates in the selection algorithm, we
implemented the algorithm presented in Sec.~\ref{sec:implementation} in
LAMMPS. To this end, we duplicated the code as \texttt{fix bond/create/random}
and the code is available online~\cite{fbc_random} under the GPL license version
2 that LAMMPS uses.
The parallel communication routines are those provided by LAMMPS for fixes, into
which we pack candidate lists for all the particles.

There is no {\em state} property in LAMMPS and the reaction is controlled by the
number of bonds. As the initiation reaction for chain growth leaves the
initiator with one bond while further reaction events leave the particles with
two bonds we define the reaction twice:
\begin{equation}
  \cee{P^{\ast} + M ->[k] P-P^{\ast}}
\end{equation}
and
\begin{equation}
  \cee{P_{n-1}-P^{\ast} + M ->[k] P_n-P^{\ast}}
\end{equation}
where $P^{\ast}$ has a different {\em type} depending on whether it is already
part of a chain or not.

\section{Simulations of chain growth}
\label{sec:sim-chain}

Simulations of chain growth start with $N_{P^\ast}$ P$^\ast$ active units
while the bulk of the simulation box is filled with monomer units M, for a total
number of particles $N=N_{P^\ast}+N_{M}=10^{4}$, the initial number fraction of polymer is
equal to the concentration of active sites
\begin{equation}
  \phi_0 = \frac{N_{P^\ast}}{N}
\end{equation}

\begin{table}[h]
  \caption{\label{tab:params}
    Parameters used for the simulations. C1 and C2 are the single and multiple
    chains growth simulations, respectively. S are the step growth simulations.
    The species A and T are integer indices corresponding to the particles type
    in ESPResSo++.
    In LAMMPS the simulation parameters are the same. Besides A and T, we denote
    by A' and T' the integer types of A and T after reaction as there is no
    state variable. The maximum number of bonds allowed given to
    \texttt{fix bond/create} are also indicated.
  }
  \begin{ruledtabular}
    \begin{tabular}{lllllllll}
      \multicolumn{9}{l}{ESPResSo++}\\
      Run & A & $\delta$A & $s_{min}^A$& T & $\delta$T & $k$ & $\Theta$ & $N_{tot}$\\
      C1  & 0 & -1        & 2         & 0 & 1         & $1$, $0.1$ and $0.01$ & 10 & $10000$\\
      C2  & 0 & -1        & 2         & 0 & 1         & $0.1$ & 25 & $10000$\\
      S   & 1 & -1        & 1         & 0 & 1         & $0.1$ and $0.01$ & 25 & 13500\\
    \end{tabular}
  \begin{ruledtabular}
  \end{ruledtabular}
    \begin{tabular}{lllllll}
      \multicolumn{7}{l}{LAMMPS}\\
                 & A &  T        & A'        & T'& A max & T max\\
      initiator  & 4 &  1        & 3         & 2 & 1 & 1\\
      propagator & 2 &  1        & 3         & 2 & 2 & 1\\
      step       & 3 &  1        & 3         & 2 & $f$ & 1
    \end{tabular}
  \end{ruledtabular}
\end{table}

The number of particles in the states M, P and P$^\ast$ is monitored for
comparison with the rate equation. The resulting polymer fraction
\begin{equation}
  \phi(t) = \frac{N_{P}+N_{P^\ast}}{N}
\end{equation}
is then plotted for a proper comparison with the figures from Akkermans {\em et
  al}~\cite{akkermans_et_al_jcp_1998}.

To obtain numerical data for $\langle N_{P^\ast M} \rangle$, for different chain
lengths, simulations of single chains are run in which the growth is stopped
when the polymer chain reaches $n$ monomers.
The integral of $g_{P^{\ast}M}(r)$ up to the cutoff radius is then used to obtain
\begin{equation}
  \langle N_{P^{\ast}M} \rangle = \int_0^{r_c} 4 \pi r^2 g_{P^{\ast}M}(r) dr ~.
\end{equation}
We observe a saturation of $\langle N_{P^{\ast}M} \rangle \approx 3.25$ with the chain length
and use this limit value to compute $k_c$.

\begin{figure}[h]
  \centering
  \includegraphics[width=\linewidth]{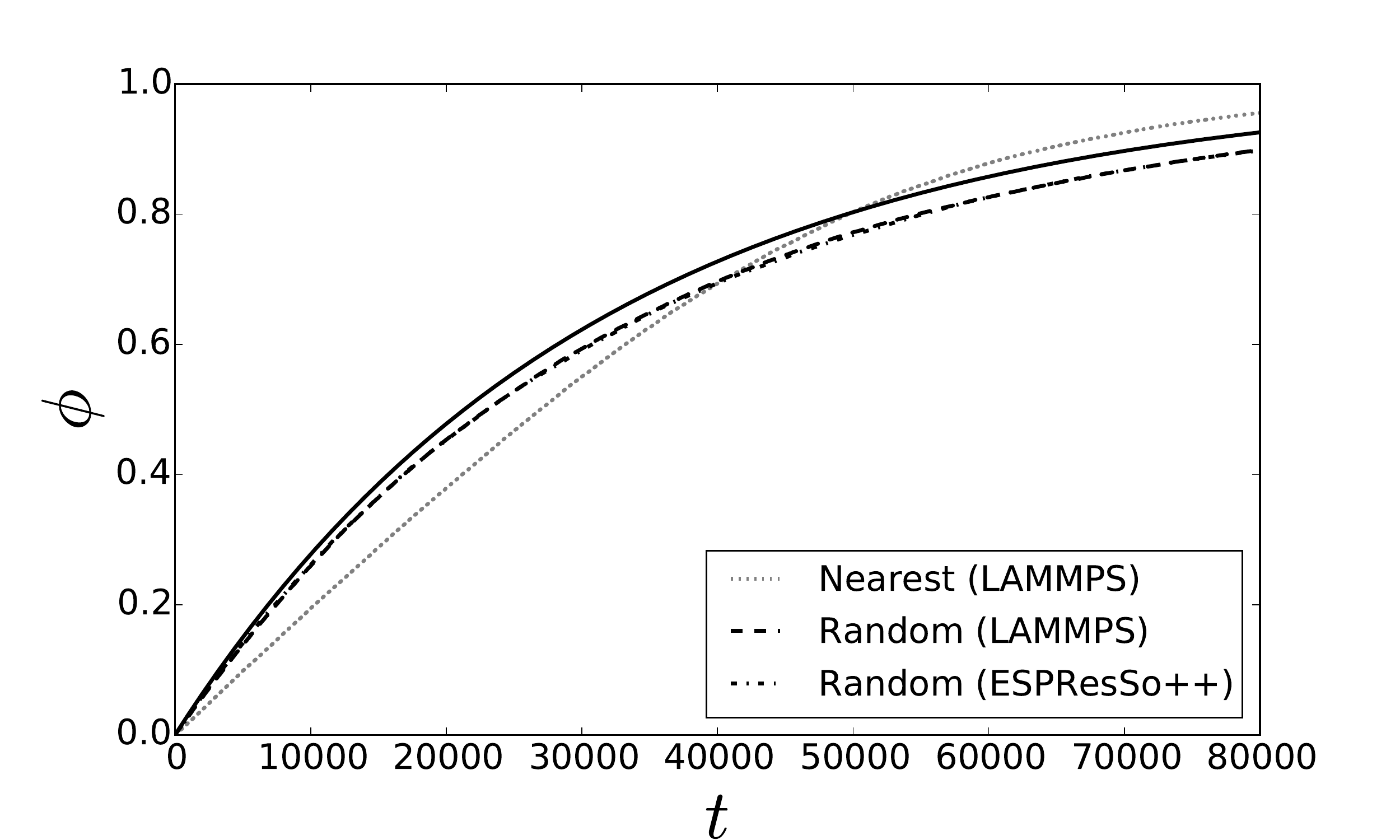}
  \caption{Simulations of chain growth for the different algorithms. The full
    line is from Eq.~\eqref{eq:chain-fraction}. The random algorithm follows the
    theory albeit a little slowlier. The implementations in ESPResSo++ and
    LAMMPS are indistinguishable. The existing algoritm in LAMMPS (``Nearest'')
    does not follow the same kinetics and could not be fitted with an
    exponential function.}
\label{fig:algos}
\end{figure}

The first round of simulations, in Fig.~\ref{fig:algos} compares the algorithm
in ESPResSo++ and in LAMMPS (existing and new).
The existing algorithm in LAMMPS that selects the nearest partners for reaction
does not follow the rate equation. To verify that difference arises from the
reaction algorithm, we have re-implemented our algorithm in LAMMPS and obtain
results that superimpose perfectly.
Further simulations with LAMMPS only use this new algorithm.

\begin{figure}[h]
  \centering
  \includegraphics[width=\linewidth]{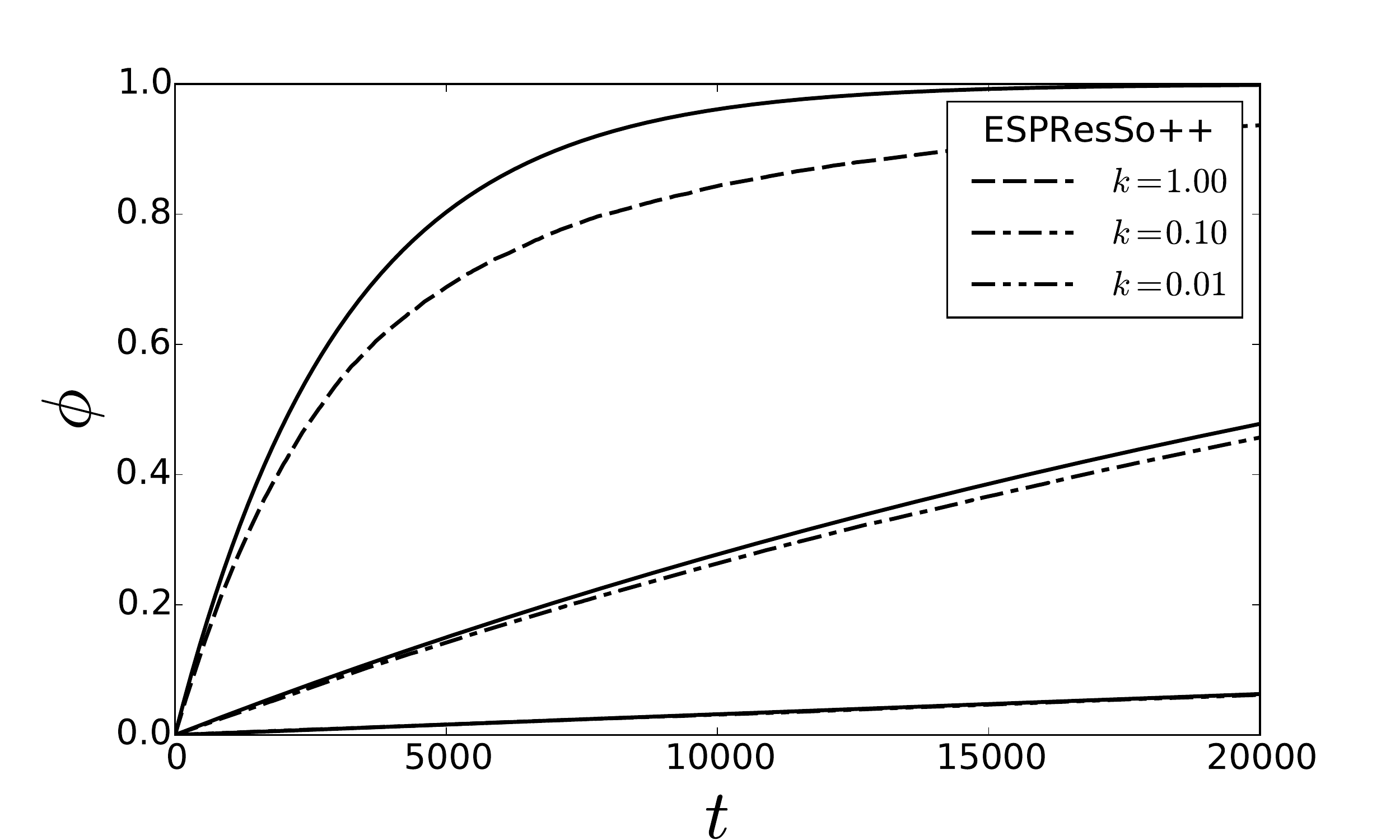}
  \includegraphics[width=\linewidth]{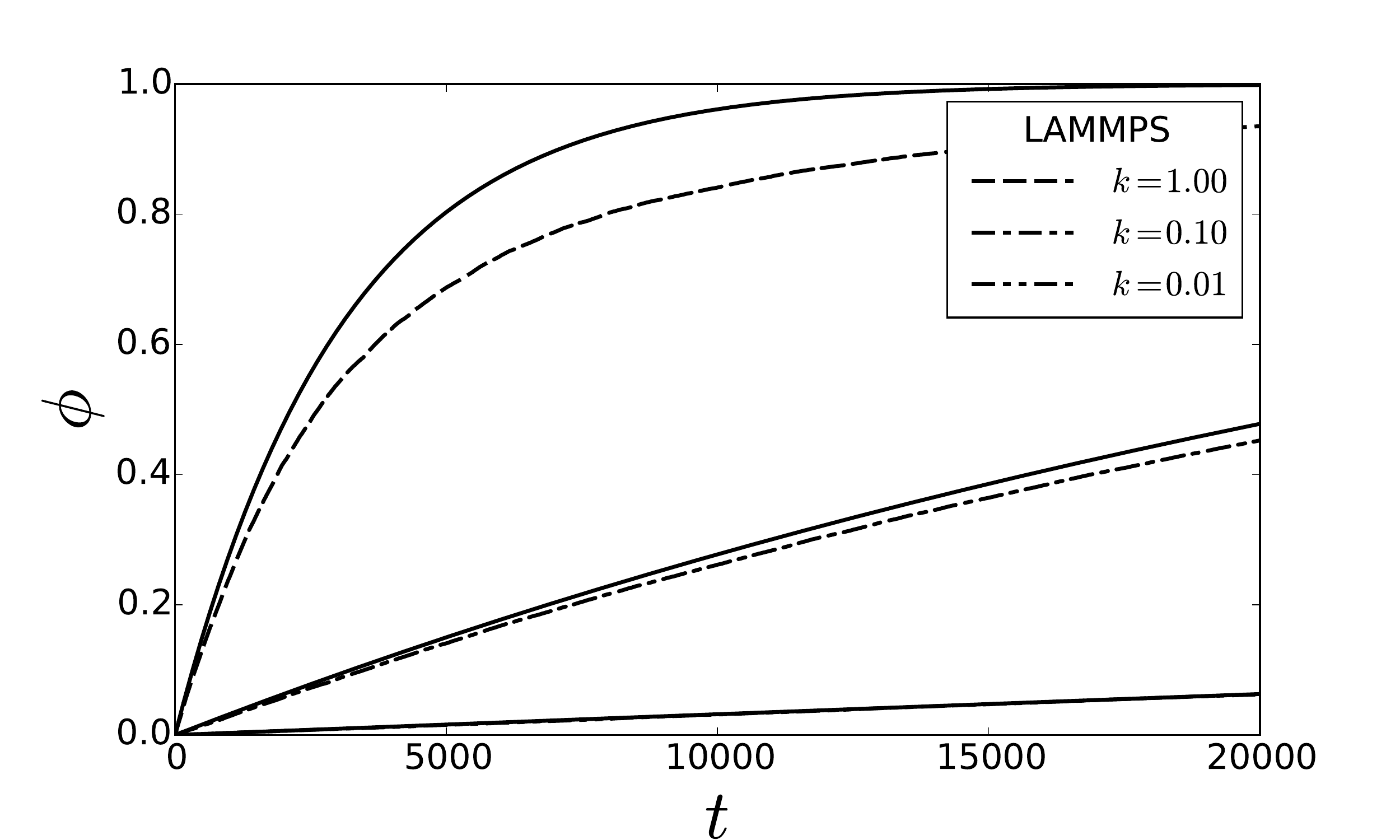}
  \caption{Simulations of chain growth (dash-dotted curves), for a single chain,
    for different intrinsic rates $k$ (see runs C1 of table~\ref{tab:params} for the parameters).
    Each curve is the average over eight
    realizations. The full lines are the theoretical estimates for the
    corresponding effective rates from Eq.~\eqref{eq:chain-fraction}.
    For the fastest rate $k=1$ the theoretical estimates overestimates the
    monomer consumption rate, with respect to the simulation results.
    The discrepancy is reduced for $k=0.1$ and for the lowest value ($k=0.01$),
    the simulation results and the theory are undistiguishable.
  }
\label{fig:single-chain}
\end{figure}

Further chain growth simulations were performed with a single chain, for different rates
$k$, and are displayed in Fig.~\ref{fig:single-chain}. As found by Akkermans
{\em et al}~\cite{akkermans_et_al_jcp_1998}, the rate equation only compares
well for low values of $k$.
When the reaction rate is too high, the active end of the chain is not given
enough time to find a new partner by molecular diffusion.

\begin{figure}[h]
  \centering
  \includegraphics[width=\linewidth]{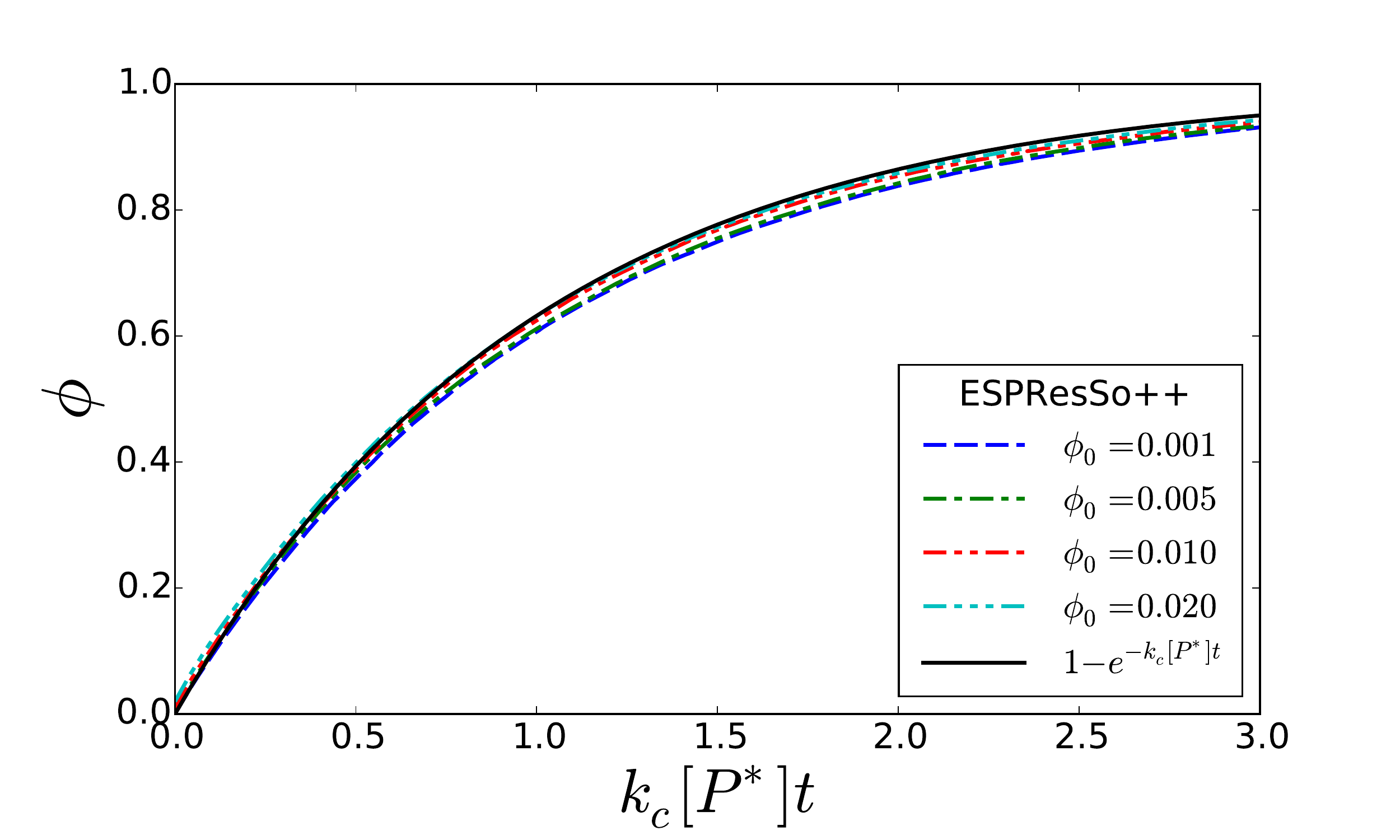}
  \includegraphics[width=\linewidth]{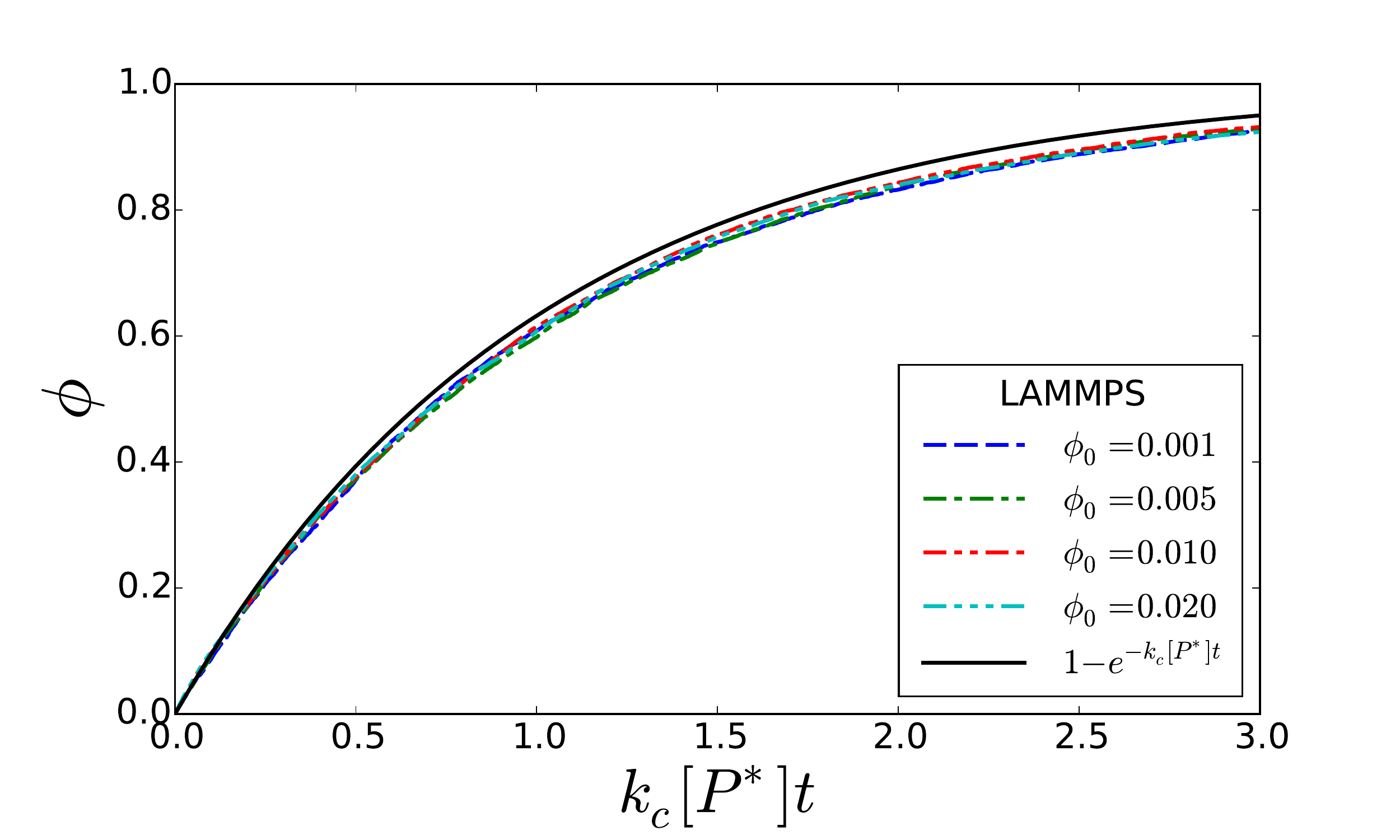}
  \caption{Simulations of chain growth for multiple chains, for initial polymer
    fractions $\phi_0=1, 3, 5, 10, 15$ and $20~10^{-3}$ and rate $k=0.1$
    (see runs C2 of table~\ref{tab:params} for the parameters).
    The time is rescaled by the effective rate $k_{c}[P^\ast]$ so that
    the theoretical estimate (full line) matches all simulation data.}
  \label{fig:several-chains}
\end{figure}

To assess the behaviour of multiple chains growth, corresponding simulations
have been run with initial polymer fractions $\phi_0$ of $1, 3, 5, 10, 15$ and
$20~10^{-3}$. The resulting $\phi(t)$ is displayed as a fraction of the scaled
time $k_c[P^\ast]t$ in Fig.~\ref{fig:several-chains}.

Given enough time, all chain growth simulations were observed to approach
$\phi=1$, similarly to the limit of Eq.~\eqref{eq:chain-fraction}.

\section{Simulations of step growth}
\label{sec:sim-step}

We have performed simulations of step growth of a model system consisting of
polymer chains $E-(P)_n-E$ with $n=3$, thus consisting of five monomer units,
and of crosslinkers $X^f$.
The simulations have been run with $f=0,1,2,3,4,5$ and rate $k=0.1$ and $0.1$
for a system of 2500 chains $E-(P)_3-E$ and 1000 crosslinkers $X$, for a total
of 13500 particles in the system.
These parameters give a stoechiometric ratio for $f=5$. They have been used for
all values of $f$ to have only a single parameter vary across the simulations.

First, the radial distribution function $g_{X^sE^0}(r)$ between the crosslinker
$X^{s\ast}$ and available end-unit $E^{0\ast}$ has been computed from
simulations with $f=0,1,2,3,4$ and $5$, and $k=0.1$, where the polymerization
runs for 5000 time units and is then stopped.
The sampling for $g_{X^sE^0}(r)$ is done for 5000 subsequent time steps.
The integral of $g_{X^sE^0}(r)$ up to the cutoff radius is then used to obtain
\begin{equation}
  \langle N_{X^sE^0} \rangle = \int_0^{r_c} 4 \pi r^2 g_{X^sE^0}(r) dr ~.
\end{equation}
The values of $\langle N_{X^sE^0} \rangle$ are given in Table~\ref{tab:nxe} for
reference.

\begin{table}[h]
  \centering
  \begin{ruledtabular}
    \begin{tabular}{llllllllll}
      $s$ & 0 & 1 & 2 & 3 & 4 & 5\\
      $\langle N_{X^sE^0} \rangle$ & 1.30 & 0.890 & 0.572 & 0.327 & 0.137 & 1.28 10$^{-2}$
    \end{tabular}
  \end{ruledtabular}
  \caption{The average number of available neighbors for curing in the
    step-growth simulations.}
  \label{tab:nxe}
\end{table}

\begin{figure}[h!]
  \centering
  \includegraphics[width=\linewidth]{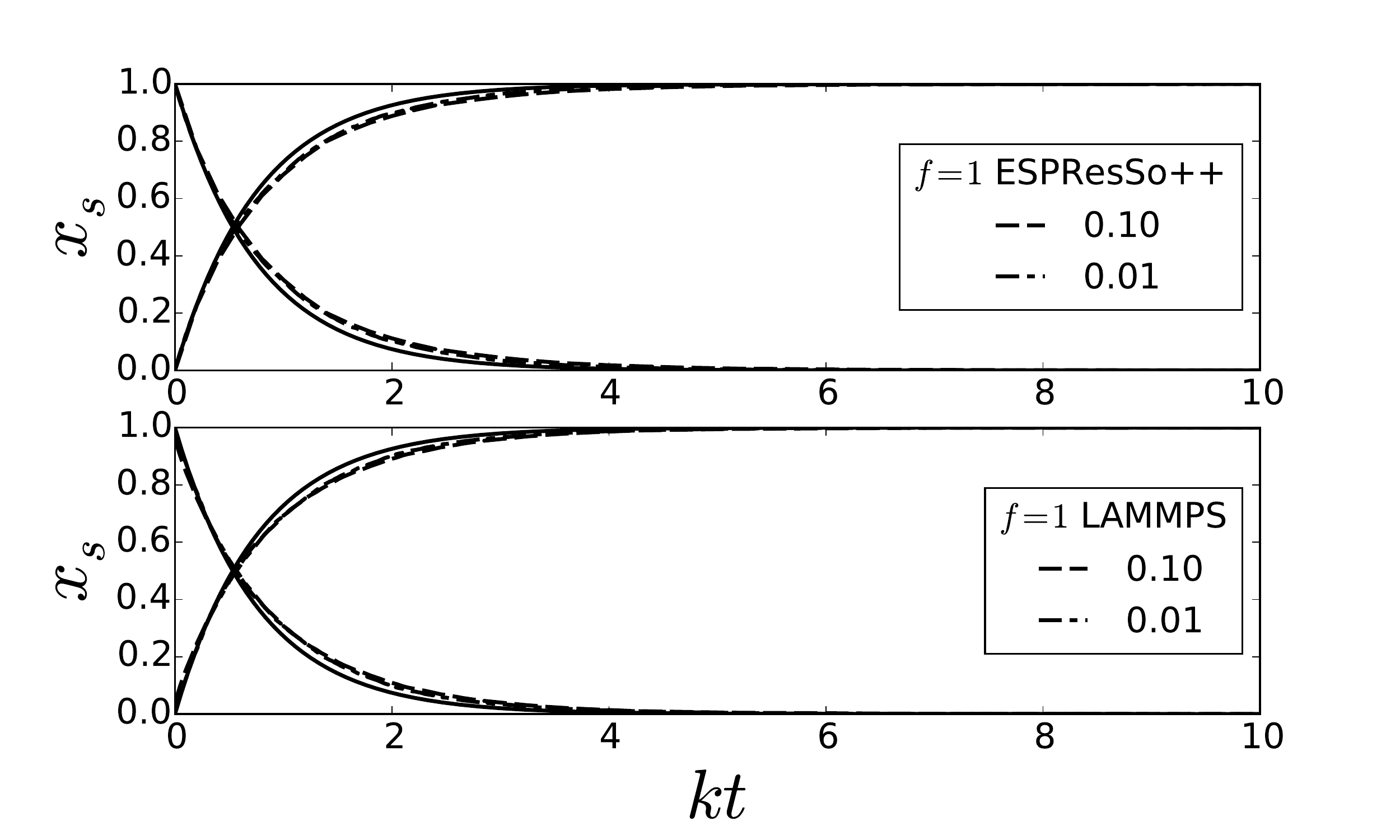}
  \includegraphics[width=\linewidth]{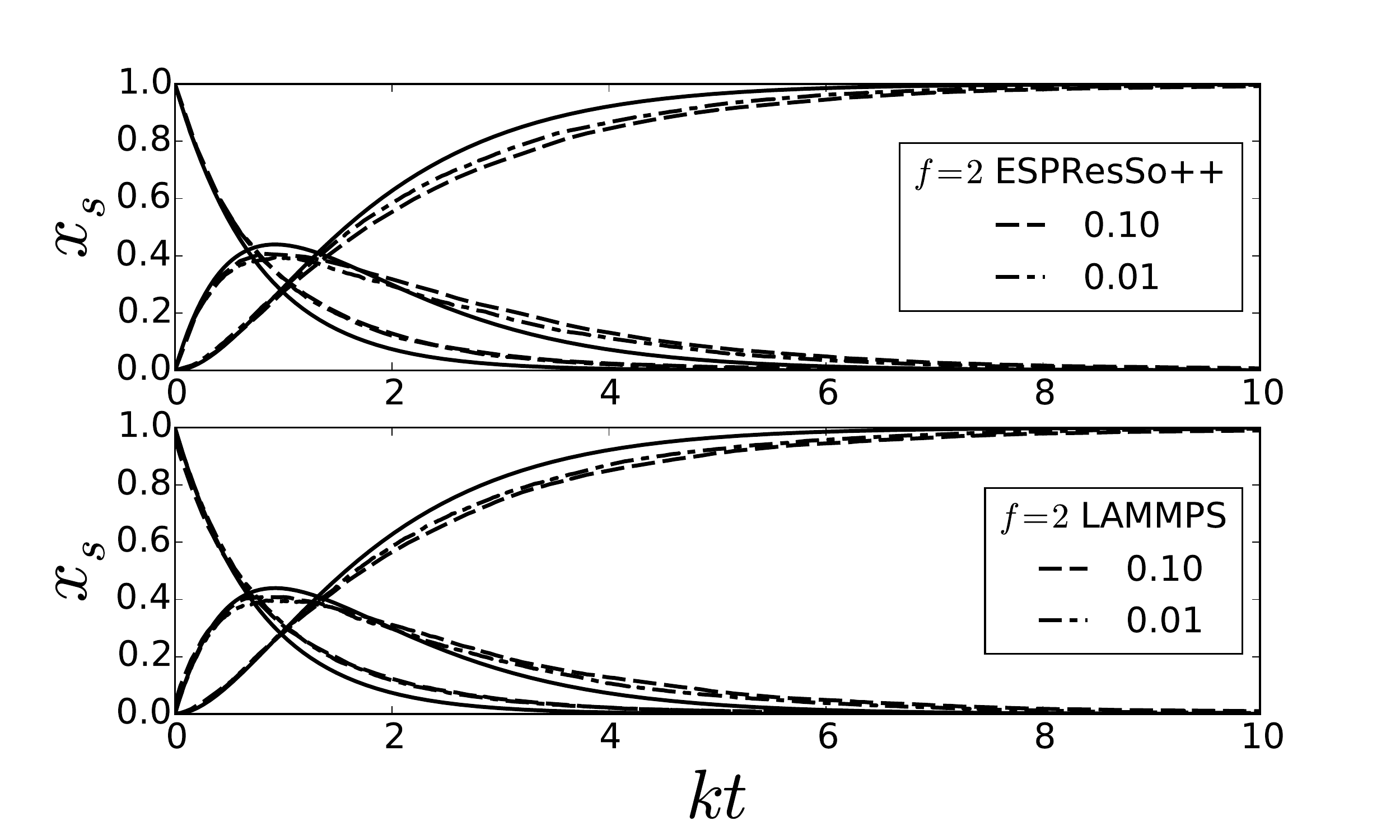}
  \includegraphics[width=\linewidth]{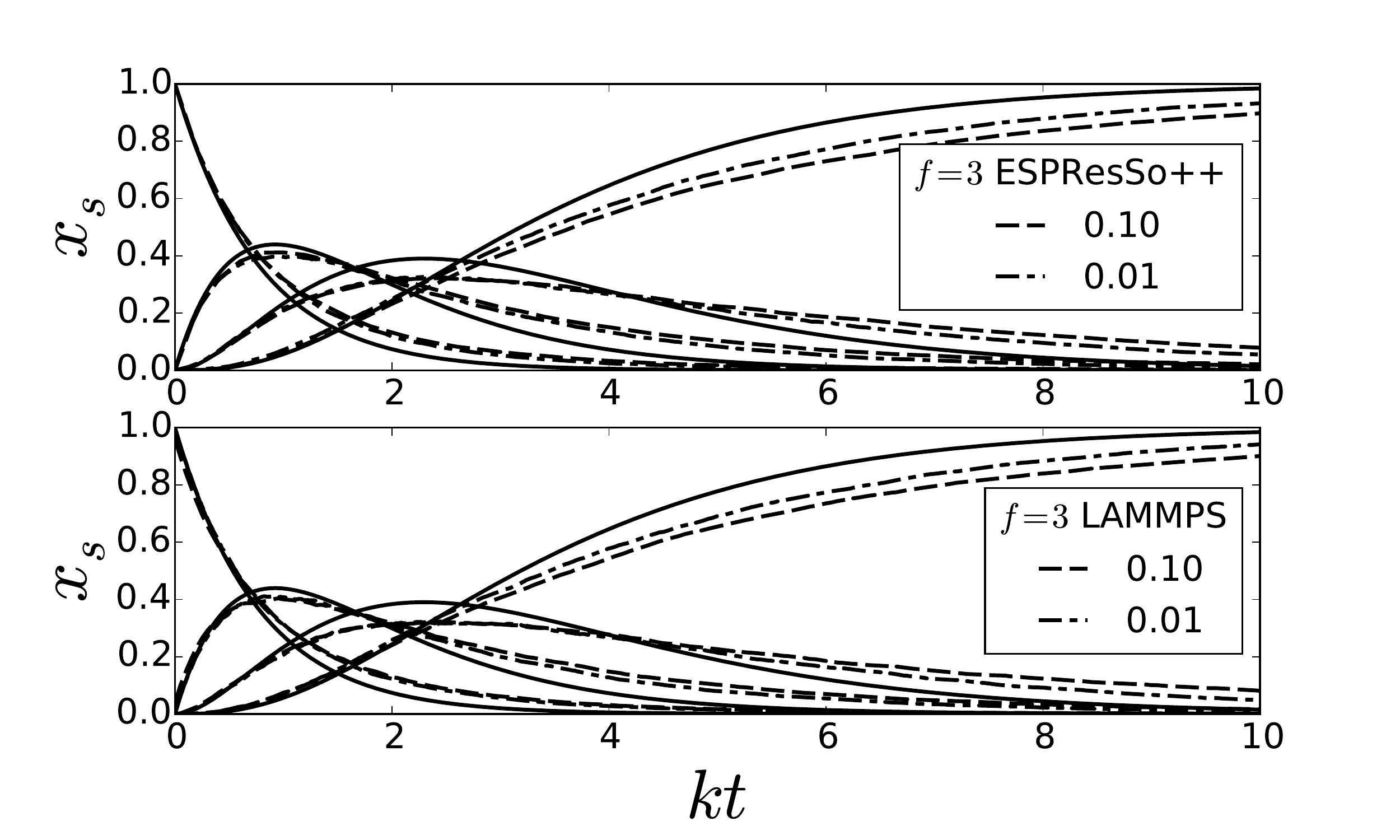}
  \caption{Concentration of crosslinker states for the step-growth simulations
    with $f=1,2$ and $3$. The full black lines are obtained from
    Eqs.~\eqref{eq:step-kin}, and the dashed and dashed dotted lines come from
    simulations with $k=0.1$ and $k=0.01$, respectively. Each simulation result
    is the average over eight realizations.
    See the parameters for runs S in table~\ref{tab:params}.
    The curves starting at $x_{s}=1$ correspond to $x_0$.
    The curves starting at $x_s$ correspond $x_s$ with $s>0$,
    where the curves growing faster initially correspond to a lower $s$
    (the fastest growing curve is for $s=1$, and so on).
    For $f=1$ and $2$, the rate equation captures properly the evolution of
    crosslinking.
    For higher functionalities, it fails to track the quantitative evolution
    beyond $kt\approx 1$ (see Fig.~\ref{fig:step-log} for $f=4$ and $5$).
  }
  \label{fig:step-all}
\end{figure}

Then, the polymerization has been studied in simulations where two rates have
been used, $k=0.1$ and $0.01$ and the results are shown in
Fig.~\ref{fig:step-all}.

\begin{figure}[h]
  \includegraphics[width=\linewidth]{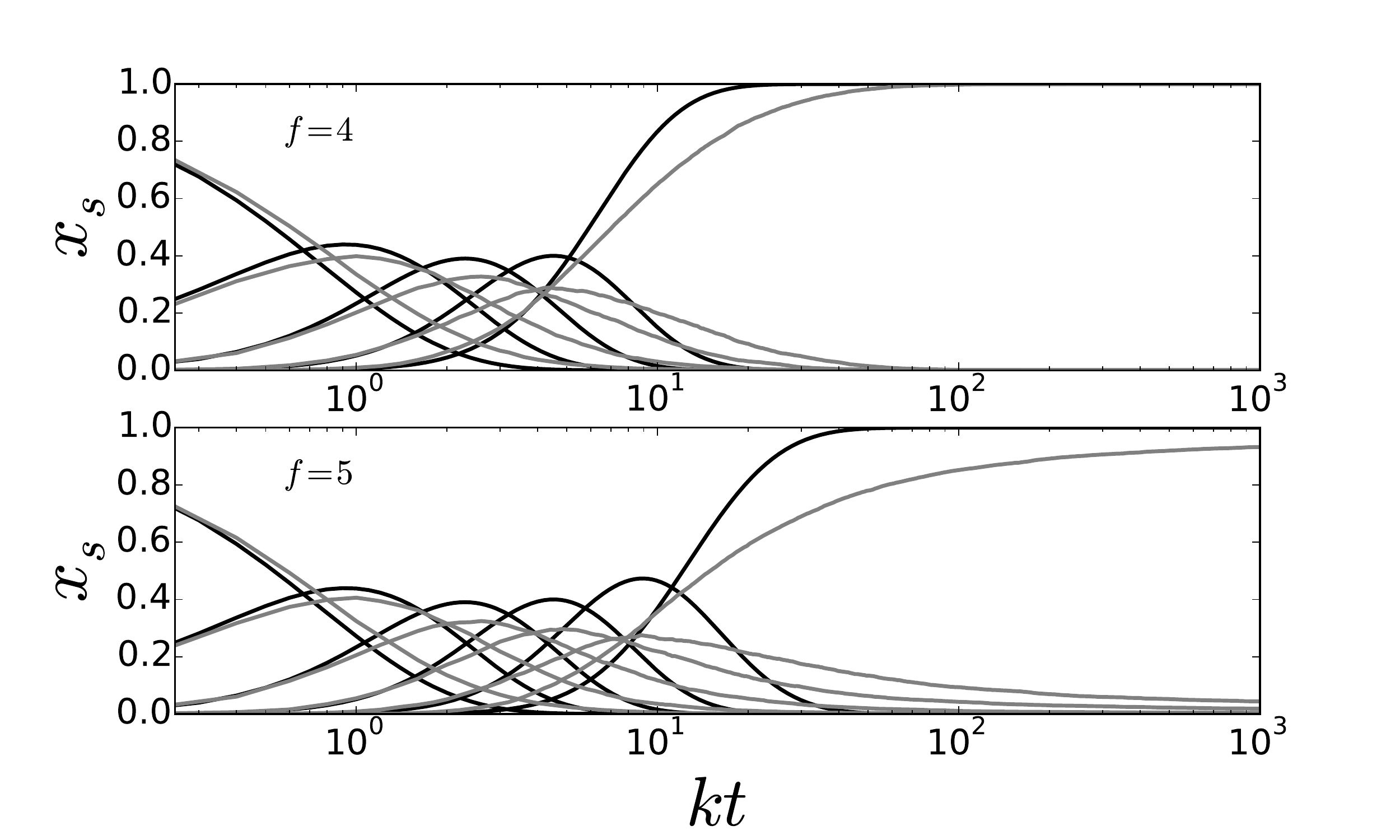}
  \caption{Same data as in Fig.~\ref{fig:step-all} for $k=0.1$ and $f=4$ and
    $5$, with a logarithmic axis for the time (x-axis).
    The full black lines are obtained from Eqs.~\eqref{eq:step-kin}, and the
    grey lines come from simulations.
    Data is shown up to the end of the simulation, to highlight the saturation
    of $x_5$ (bottom panel) below the limiting value of $1$ that is
    reached for $f=4$ (top panel).
    Simulations performed with ESPResSo++.}
  \label{fig:step-log}
\end{figure}

For low functionality ($f=1$ or $2$), the concentrations $[X^{s\ast}]$ given by
Eqs.~\eqref{eq:step-kin} compare well to the ones from the simulations.  The
simulation data shows a delay in the polymerization process, with respect to the
rate equation, similarly to what has been observed for chain growth (see
Sec.~\ref{sec:sim-chain}).
For higher functionality, the rate equation compares well to the simulation data
only for the initial stages of the polymerization. Results for $kt$ up to $1.5$
are displayed to highlight the proper capture of the initial polymerization
kinetics.

The discrepancy between the rate equation and the simulation data is
unavoidable, as the rate equation only considers the average value for the
number of reaction candidates, and highlights a motivation to develop the full
simulation model.

Figure~\ref{fig:step-log} presents the same data as Fig.~\ref{fig:step-all} with
a larger time span (for $f=4$ and $f=5$ only).
For $f=4$, the fraction $x_4$ of fully crosslinked X particles
saturates at 1 (maximum value), similarly to the kinetic model.
For $f=5$, besides the observed lag in the polymerization, we observe that
$x_5$ does not reach the same saturation value.
Indeed, crosslinkers having already formed four bonds (in the case $f=5$) have
on average $0.01$ neighbours.
This average hides the fact that many of these crosslinkers have zero neighbors
of type and state E$^{0\ast}$ that would allow further reaction.
The polymerization is thus stopped by an effective depletion of reactant.

\section{Conclusions}

We have presented an adaptable algorithm for thermoneutral polymerization in parallel
Molecular Dynamics (MD) simulations.
The algorithm handles several polymerization mechanisms and may involve
molecular compounds in which only selected sites participate in the
polymerization process, as was done here for step growth.
A difference in performance between ESPResSo++ and LAMMPS is observed,
consistently with the observations made by the developers of
ESPResSo++\cite{halverson_et_al_espressopp_cpc_2013}.
Other criteria should guide the choice of the simulation package: the type of
model simulated or the use of the Python interface, for instance.

The kinetic model of Akkermans {\em et al}~\cite{akkermans_et_al_jcp_1998} was
validated on the chain growth results and a kinetic model for step growth was
introduced and compared favorably to the simulations.
A systematic delay of the simulation process in the simulation is found for both
growth mechanisms. That delay was also found in
Refs.~\onlinecite{akkermans_et_al_jcp_1998,hoy_fredrickson_jcp_2009} and is
caused by the simplifications made in the rate equations with respect to the
full molecular simulations.
The polymerization algorithm has been implemented in ESPResSo++ and compared to
the corresponding feature of LAMMPS. As a different kinetic evolution was found,
we proceeded to implement our algorithm in LAMMPS to verify that this would
bring the results in agreement, which was the case for chain growth and for step
growth.

Due to the relative simplicity of coarse-grained models, with respect to
atomistic descriptions, it is possible to control the polymerization process in
its time evolution and to avoid typical artifacts such as energy jumps and
discontinuous trajectories.
On the basis of the present work, it is possible to backmap a system's
coordinates to the atomistic level after the polymerization process.
Several extensions of the algorithm are feasible: introduce several
concurrent chemical reactions with different intrinsic rates or further
constrain the reaction acceptance to conformation properties (e.g. to avoid
unrealistic angles in the newly formed molecule).

While the present work is limited to irreversible reactions, other works have
already considered interaction potentials than ``break'' past a given
cutoff~\cite{tsige_stevens_macromolecules_2004}.
An alternative approach to the dissociation process is to consider a stochastic
rate at which a bond dissapears~\cite{caby_et_al_living_filaments_jcp_2012}.
This latter approach does not achieve energy conservation however.
No solution that combines continuous trajectories and stochastic dissociation
has been proposed yet.

\begin{acknowledgments}
The authors acknowledge fruitful interactions with the developers of ESPResSo++.
This work was supported by the ``Strategic Initiative Materials'' in Flanders
(SIM) under the InterPoCo program.
The computational resources and services used in this work were provided by the
VSC (Flemish Supercomputer Center), funded by the Hercules Foundation and the
Flemish Government -- department EWI.
\end{acknowledgments}

\appendix

\section{Simulation protocols in ESPResSo++ and LAMMPS}
\label{sec:protocols}

ESPResSo++ simulation protocol for chain and step growth:
\begin{enumerate}
\item Place particles at random in the simulation box. Chains for the step
  growth simulations are placed ``one chain at a time'' using the random-walk
  placement routine {\tt espresso.tools.topology.polymerRW} of ESPResSo++.
\item Enable the velocity rescaling thermostat.
\item Run a warmup integration in which the interaction potential are capped at
  a maximum value.
\item Run a warmup integration in which the interaction potential are uncapped.
\item Disable the thermostat.
\item Run the ``production'' run, with the polymerization mechanism enabled.
\end{enumerate}

LAMMPS simulation protocol for chain growth:
\begin{enumerate}
\item Place particles at random in the simulation box.
\item Enable the \texttt{temp/rescale} thermostat and use the \texttt{nve/limit}
  displacement limiter.
\item Run a warmup integration.
\item Disable the thermostat and displacement limiter.
\item Run the ``production'' run, with the polymerization mechanism enabled.
\end{enumerate}

LAMMPS simulation protocol for step growth:
\begin{enumerate}
\item Replicate regularly a single chain in a low-density simulation box.
\item Place crosslinkers at random in the simulation box.
\item Enable the \texttt{temp/rescale} thermostat and use the \texttt{nve/limit}
  displacement limiter.
\item Iterate over MD runs and minimization steps.
\item Increase gradually the density to the target value with \texttt{fix deform}.
\item Disable the thermostat and displacement limiter.
\item Run a warmup integration with the \texttt{nvt} thermostat (Nosé-Hoover) at
  the target temperature.
\item Disable the thermostat.
\item Run the ``production'' run, with the polymerization mechanism enabled.
\end{enumerate}

\bibliographystyle{apsrev4-1}
\bibliography{parallel_polym}

\end{document}